# Compact Modeling of the Effects of Parasitic Internal Fringe Capacitance on the Threshold Voltage of High-K Gate Dielectric Nanoscale SOI MOSFETs

M. Jagadesh Kumar[1] *Senior Member, IEEE,* Sumeet Kumar Gupta and Vivek Venkataraman

Department of Electrical Engineering,

Indian Institute of Technology, Delhi,

Hauz Khas, New Delhi – 110 016, INDIA.

Email: mamidala@ieee.org  Fax: 91-11-2658 1264

[1]Corresponding author

**Abstract**

A compact model for the effect of parasitic internal fringe capacitance on threshold voltage in high-K gate dielectric SOI MOSFETs is developed. Our model includes the effects of the gate dielectric permittivity, spacer oxide permittivity, spacer width, gate length and width of MOS structure. A simple expression for parasitic internal fringe capacitance from the bottom edge of the gate electrode is obtained and the charges induced in the source and drain regions due to this capacitance are considered. We demonstrate an increase in surface potential along the channel due to these charges resulting in a decrease in the threshold voltage with increase in gate dielectric permittivity. The accuracy of the results obtained using our analytical model is verified using 2-D device simulations.

# 1. Introduction

The gate oxide thickness in modern small geometry MOSFETs is approaching the tunneling limit for electrons, resulting in an increase in gate leakage current. Scaling the effective gate dielectric thickness will require alternative materials, with higher permittivities ($\varepsilon_{ox}$) and greater physical thicknesses (a factor of $\varepsilon_{ox}/\varepsilon_{SiO_2}$) to prevent direct gate tunneling. However, the use of a high-K gate material may result in dielectric thicknesses comparable to the device gate length, resulting in increased fringing fields from the gate to the source/drain regions compromising the short-channel performance [1,2].

Kamchouchi and Zaky [3] developed a model for the parasitic capacitance associated with the bottom edge of the gate electrode in which it is assumed that (i) the dielectric is the same throughout and (ii) its thickness ($t_{ox}$) is much less than the gate length ($L_g$). However, in scaled down MOSFETs, the gate dielectric and the spacer oxide have different permittivities. In this paper, we develop a simple expression for the internal fringe capacitance ($C_{bottom}$) considering different gate and spacer dielectric constants ($\varepsilon_{ox} \neq \varepsilon_{sp}$) and gate dielectric thickness comparable to the gate length.

Young [4] developed a model for the surface potential along the channel for SOI MOSFETs without considering the effect of the internal fringe capacitance ($C_{bottom}$). In this paper, we have demonstrated the effect of $C_{bottom}$ on the surface potential by considering the charges induced on the source and drain regions (due to fringing field lines from the bottom of the gate electrode) and the potential developed due to these charges in the channel region.

Kumar and Chaudhry [5] and Reddy and Kumar [6] have earlier developed a model for the threshold voltage ($V_{th}$) for Dual Material Gate SOI MOSFETs. Using a similar approach, we have developed a model for the threshold voltage for the Single Material Gate SOI MOSFETs by including the effect of internal fringe capacitance on threshold voltage, which can be easily solved using a few iterations. This model thus provides an efficient tool for design and characterization of high-K gate dielectric SOI MOSFETs including the effects of parasitic internal fringe capacitance. The effects of varying device parameters can easily be investigated using the simple models presented in this work. The model results are verified by comparing them with the 2-D simulated results from MEDICI [10].

## 2. Model for internal fringe capacitance ($C_{bottom}$)

A schematic cross-sectional view of a SOI MOSFET with high-K gate dielectric is shown in Fig.1 with the fringing field lines from the bottom of the gate electrode to the drain and source regions. For simplicity, we assume circular field lines as shown in Fig.2, similar to the approach used in earlier works for the other components of parasitic capacitance [7,8,9]. The infinitesimal capacitances in the high-K and spacer regions, respectively, can be written as

$$dC_1 = \frac{\varepsilon_{ox} W dy}{(\pi/2)y} \qquad \text{and} \qquad dC_2 = \frac{\varepsilon_{sp} W dy}{(\pi/2)(t_{ox} - y)}$$

where $\varepsilon_{ox}$ is the permittivity of gate dielectric , $\varepsilon_{sp}$ is the permittivity of spacer material, $W$ is width of MOS structure and $t_{ox}$ is the gate dielectric thickness. Since $dC_1$ and $dC_2$ are in series, the net infinitesimal capacitance can be written as

$$dC = \frac{dC_1 dC_2}{dC_1 + dC_2} = \frac{2\varepsilon_{ox}\varepsilon_{sp} W dy}{\pi(\varepsilon_{ox}t_{ox} + \varepsilon_{sp}y - \varepsilon_{ox}y)} \tag{1}$$

The total internal fringe capacitance can be obtained by integrating (1) over the gate dielectric thickness as

$$C_{bottom} = \int_0^{t_{ox}} \frac{2\varepsilon_{ox}\varepsilon_{sp}Wdy}{\pi(\varepsilon_{ox}t_{ox} + \varepsilon_{sp}y - \varepsilon_{ox}y)} = \frac{2\varepsilon_{ox}\varepsilon_{sp}W}{\pi(\varepsilon_{ox} - \varepsilon_{sp})}\ln\left(\frac{\varepsilon_{ox}}{\varepsilon_{sp}}\right) \tag{2}$$

Equation (2) is refined below in order to take care of the imperfect circularity of the fringing field lines.

The model by Kamchouchi and Zaky [3] gives the parasitic capacitance per unit length considering the electric field lines fringing from the entire perimeter of the bottom edge of the gate electrode as

$$C = \frac{(2 - \ln 4)\varepsilon_{ox}}{2\pi} \qquad \text{for} \qquad \varepsilon_{ox} = \varepsilon_{sp} \text{ and } \frac{t_{ox}}{L_g} << 1 \tag{3}$$

where $L_g$ is gate length. The total fringe capacitance can be obtained by multiplying (3) with the perimeter of the bottom edge of the gate electrode [3]. Since we need to account for electric field lines fringing from the bottom edge of the gate to either the source or the drain region only, the internal fringe capacitance can be written as

$$C_{bottom} = \frac{(2 - \ln 4)\varepsilon_{ox}W}{2\pi} \cong \frac{(0.3)\varepsilon_{ox}W}{\pi} \qquad \text{for} \qquad \varepsilon_{ox} = \varepsilon_{sp} \text{ and } \frac{t_{ox}}{L_g} << 1 \tag{4}$$

It can be seen that equation (2) reduces to equation (4) in the limit $\varepsilon_{ox} \rightarrow \varepsilon_{sp}$ but for a constant factor. This difference is arising because of the assumption of the fringing electric field lines being circular while deriving equation (2). Thus, equation (2) is multiplied by the above factor (0.3/2 = 0.15) to obtain

$$C_{bottom} = \frac{(0.3)\varepsilon_{ox}\varepsilon_{sp}W}{\pi(\varepsilon_{ox} - \varepsilon_{sp})}\ln\left(\frac{\varepsilon_{ox}}{\varepsilon_{sp}}\right) \tag{5}$$

The above expression reduces to (4) in the limit $\varepsilon_{ox} \rightarrow \varepsilon_{sp}$.

The Kamchouchi and Zaky model in equation (4), and hence (5), assumes that the separation between the electrodes is very small in comparison to the length of the electrodes. This is certainly not the case in short channel high-K dielectric SOI MOSFETs. Hence (5) is further modified to represent the true picture.

The fringing field from the gate to the source/drain regions increases as a function of $\frac{t_{ox}}{L_g}$[1]. This is the effect of increased crowding of field lines in the spacer region for large gate dielectric thicknesses, i.e for $t_{ox}$ comparable to $L_g$. Hence, the fringe capacitance in the spacer region is more than what has been assumed while deriving (5). To account for this, an effective spacer dielectric constant is defined as

$$\varepsilon'_{sp} = \left(1 + \frac{t_{ox}}{L_g}\right)\varepsilon_{sp} \tag{6}$$

Substituting (6) in place of $\varepsilon_{sp}$ in (5), we can obtain the final expression for the parasitic internal fringe capacitance as

$$C_{bottom} = \frac{(0.3)\varepsilon_{eff}W}{\pi} \tag{7}$$

where

$$\varepsilon_{eff} = \frac{\varepsilon_{ox}\varepsilon'_{sp}}{\varepsilon_{ox} - \varepsilon'_{sp}}\ln\left(\frac{\varepsilon_{ox}}{\varepsilon'_{sp}}\right)$$

## 3. Effect of parasitic internal fringe capacitance on surface potential

To include the effect of internal fringe capacitance on surface potential, we assume the charge distribution induced in the source and drain regions due to the fringing electric field lines as in a uniformly charged plate as shown in Fig. 3 with the charge density σ given by

$$\sigma = \left( \frac{C_{bottom} V_p}{W.t} \right) \qquad \text{with} \qquad \begin{aligned} V_p &= V_{bi} - V_G + V_{FB} \qquad \text{for source region} \\ V_p &= V_{bi} + V_D - V_G + V_{FB} \quad \text{for drain region} \end{aligned}$$

where $V_{bi} = (E_g/2) + V_T ln(N_A/n_i)$ is the built-in potential across the body-source junction, $E_g$ is the silicon bandgap, $N_A$ is the body/substrate doping concentration, $V_T$ is the thermal voltage, $n_i$ is the intrinsic carrier concentration, $V_G$ and $V_D$ are the potentials applied to gate and drain electrodes respectively, $V_{FB}$ is the flat band voltage, $W$ is width of the gate and $t$ is the spacer thickness.

The potential due to this uniform charge distribution is evaluated along the channel at the middle of the gate width W as the effect is maximum here. In a MOSFET, field lines originating from the bottom of the charged plate only will contribute to the potential in the channel and therefore only half the coulomb potential due to these charges is considered. Then the potential at a distance 'x' from the uniformly charged plate is given by (Fig. 3)

$$V(x) = \frac{1}{2} \int_0^t \int_{-W/2}^{W/2} \frac{\sigma \, dx_1 \, dx_2}{4\pi\varepsilon_{Si}\sqrt{(x+x_1)^2 + (x_2)^2}} \tag{8}$$

where $\varepsilon_{Si}$ is the dielectric constant of silicon . Evaluation of the integral in (8) gives

$$V(x) = \frac{\sigma}{8\pi\varepsilon_{Si}} \left[ (x+t).ln\left\{ \frac{\sqrt{(x+t)^2 + (W^2/4)} + (W/2)}{\sqrt{(x+t)^2 + (W^2/4)} - (W/2)} \right\} - x.ln\left\{ \frac{\sqrt{x^2 + (W^2/4)} + (W/2)}{\sqrt{x^2 + (W^2/4)} - (W/2)} \right\} + W.ln\left\{ \frac{\sqrt{x^2 + (W^2/4)} + t + \sqrt{x^2 + t^2 + 2t\sqrt{x^2 + (W^2/4)}}}{\sqrt{x^2 + (W^2/4)} + x} \right\} \right]$$

$$\tag{9}$$

Young[4] proposed a model for the surface potential along the channel (x-direction in Fig.1) for a fully depleted SOI MOSFET assuming a simple parabolic potential profile in the vertical direction ( y-direction in Fig. 1) as

$$\phi_s(x) = A \exp(\lambda x) + B \exp(-\lambda x) - \delta \tag{10}$$

where

$$\delta = \frac{q N_A t_{Si} t_{ox}}{\varepsilon_{ox}} - V_G + V_{FB}$$

and $\lambda = \sqrt{\varepsilon_{ox}/(t_{ox} \varepsilon_{Si} t_{Si})}$, $V_{FB}$ is the flat band voltage, $t_{Si}$ is the silicon substrate thickness and $\varepsilon_{Si}$ is the permittivity of silicon. The parameters $A$ and $B$ are given as

$$A = \left\{ \frac{(V_{bi} + \delta + V_D) - (V_{bi} + \delta) \exp(-\lambda L_g)}{1 - \exp(-2\lambda L_g)} \right\} \exp(-\lambda L_g) \qquad B = \left\{ \frac{(V_{bi} + \delta) - (V_{bi} + \delta + V_D) \exp(-\lambda L_g)}{1 - \exp(-2\lambda L_g)} \right\}$$

In the above analysis, we have assumed the buried oxide thickness to be large and hence neglected the corresponding capacitance.

The surface potential minimum is at

$$x = x_{min} = \frac{1}{2\lambda} \ln\left(\frac{B}{A}\right)$$

and is given by

$$\phi_{s\,min} = 2\sqrt{AB} - \delta \tag{11}$$

To obtain the surface potential including the effect of the internal fringe capacitance, the potential due to the charges on the source and drain regions as given by (9) is added to (10) and the expression for surface potential is modified as

$$\phi_s^{'}(x) = A \exp(\lambda x) + B \exp(-\lambda x) - \delta + \left( V(x)\big|_{source} + V(L_g - x)\big|_{drain} \right) \tag{12}$$

Here $V(x)\big|_{source}$ is the potential along the channel due to the charges in the source region, and $V(L_g - x)\big|_{drain}$ is the potential along the channel due to the charges in the drain region.

The minimum of the modified surface potential is given by

$$\phi_{s\,min}' = 2\sqrt{AB} - \delta + \left( V(x)\big|_{source} + V(L_g - x)\big|_{drain} \right)\bigg|_{x = x_{min}} \tag{13}$$

## 4. Effect of parasitic internal fringe capacitance on threshold voltage

In [5], Kumar and Chaudhry proposed a model for the threshold voltage of DMG-SOI MOSFETs by equating surface potential minimum given by (11) to twice the Fermi potential i.e.

$$\phi_{s\min} = 2\phi_F \qquad \text{where} \qquad \phi_F = V_T \ln\left(N_A/n_i\right) \qquad (14)$$

The model in [5] can be modified for Single Material Gate (SMG) SOI MOSFETs as

$$V_{th} = \frac{\left(-V_{\phi 1} + \sqrt{V_{\phi 1}^2 - 4\xi V_{\phi 2}}\right)}{2\xi} \qquad (15)$$

where
$$\xi = 2\cosh\left(\lambda L_g\right) - 2 - \sinh^2\left(\lambda L_g\right)$$
$$V_{\phi 1} = V_{bi1}\left(1 - \exp\left(\lambda L_g\right)\right) + \left(4\phi_F - 2u\right)\sinh^2\left(\lambda L_g\right) - V_{bi2}\left(1 - \exp\left(-\lambda L_g\right)\right)$$
$$V_{\phi 2} = V_{bi1}V_{bi2} - \left(2\phi_F - u\right)^2 \sinh^2\left(\lambda L_g\right)$$
$$V_{bi1} = \left(V_{bi} - u\right)\left(1 - \exp\left(-\lambda L_g\right)\right) + V_{DS}$$
$$V_{bi2} = \left(V_{bi} - u\right)\left(\exp\left(\lambda L_g\right) - 1\right) - V_{DS}$$
$$u = -\frac{qN_A t_{Si} t_{ox}}{\varepsilon_{ox}} - V_{FB}$$
$$\lambda = \sqrt{\varepsilon_{ox}/\left(t_{ox}\varepsilon_{Si}t_{Si}\right)}$$

To incorporate the effect of internal fringe capacitance, we modify (14) as

$$\phi'_{s\min} = 2\phi_F \qquad (16)$$

where $\phi'_{s\min}$ is as given in (13). Solving (16) results in the same expression for the threshold voltage as given in (15) except for a modification in the expression for '$u$' given as

$$u = -\frac{qN_A t_{Si} t_{ox}}{\varepsilon_{ox}} - V_{FB} + \left(V(x)\big|_{source} + V(L_g - x)\big|_{drain}\right)\bigg|_{x = x_{\min}} \qquad (17)$$

Since $V(x)$ is dependent on $V_G$ (see (9)) and $V_{th}$ is the value of $V_G$ at $\phi'_{s\min} = 2\phi_F$, the value of threshold voltage $V_{th}$ can be easily obtained by iteratively solving (15).

## 5. Simulation Results and Discussion

Figure 4 shows the variation of internal fringe capacitance ($C_{bottom}$) versus gate dielectric permittivity evaluated using the proposed model and compared with MEDICI [10] simulations for channel lengths of 60 nm and 40 nm. The capacitance $C_{bottom}$ is extracted from MEDICI in the following manner: (i) the gate, source and drain electrode heights are made negligible so that the other components of the fringe capacitance that arise due to the finite electrode thickness are nullified, and (ii) the total capacitance as seen from the gate electrode for this MOSFET structure is extracted using the method of incremental charge due to small increment in voltage ($\partial Q/\partial V$). This gives the fringe capacitance plus the MOS gate capacitance ($C_{ox}$). Now $C_{ox}$ is subtracted from the capacitance obtained above to find the value of $C_{bottom}$. There is a difference of about $2 \times 10^{-18} F$ between our calculated and simulated values. The values of the ideal MOS gate capacitance $C_{ox} = (\varepsilon_{ox}/t_{ox})L_g W$ are $10^{-15}$ F for $L_g$ = 60 nm and $7 \times 10^{-16}$ F for $L_g$ = 40 nm . It can be seen in the figure that the parasitic fringe capacitance increases with increasing gate dielectric permittivity. This is because the physical gate dielectric thickness increases as gate dielectric permittivity increases (by a factor of $\varepsilon_{ox}/\varepsilon_{SiO_2}$) for the same EOT. This results in an increase in the fringing electric field lines from the bottom of the gate electrode to the source and drain regions.

In Figure 5, the values of surface potential with and without the effect of parasitic fringe capacitance are plotted against the horizontal distance $x$ in the channel for gate dielectric permittivity of 60. It is evident in the figure that the surface potential increases due to the effect of fringe capacitance resulting in an increase of minimum surface potential. Also it can be seen that the effect of this potential is maximum roughly at the point of minimum surface potential and hence one would expect a significant effect of $C_{bottom}$ on the threshold voltage.

To verify the proposed model, the 2-D device simulator MEDICI [10] was used to simulate the threshold voltage. A fully depleted (FD) n-channel SOI structure is implemented in MEDICI having uniformly doped source/drain and body regions. This structure is simulated both with and without gate-source/drain overlap.

In Figure 6, the calculated values of threshold voltage as a function of gate dielectric permittivity are compared with those obtained from 2-D simulation. As can be seen from the figure, the threshold voltage obtained from the model tracks the simulation values well with a maximum offset of about 15 mV. The model gives good agreement with simulation even for the case of gate–S/D overlap. It is evident that the threshold voltage decreases with increasing $\varepsilon_{ox}$. This is because of the increase in surface potential as a result of the charges induced in the drain and source regions due to fringing field lines from the bottom of the gate electrode. This results in an early onset of inversion ($\phi_{s\min} = 2\phi_F$) in the channel and hence a lower threshold voltage. The drop in threshold voltage is as high as about 60-80 mV for $L_g$=40 nm as $\varepsilon_{ox}$ increases from 3.9 to 80.

To investigate the significance of Quantum Mechanical effects in the above analysis, the simulations were performed by including QM effects also in MEDICI [10]. In Figure 7, the threshold voltage obtained with and without QM effects are compared. As is evident from the figure, there is a very small and insignificant difference between the two suggesting that QM effects can be neglected in threshold voltage calculations.

## 6. Conclusions

We have examined the effect of the parasitic internal fringe capacitance on the threshold voltage of fully depleted high-K SOI MOSFETs by developing a simple model for the internal fringe capacitance and obtaining expressions for surface potential and threshold voltage including the effect of the internal fringe capacitance. We have compared the results with accurate two-dimensional simulations. The calculated values of the threshold voltage obtained from the proposed model agree well with the simulated results. There is a significant drop in threshold voltage due to fringing field lines from the bottom edge of the gate electrode to the source and drain regions for higher gate dielectric permittivities (i.e. higher physical gate dielectric thickness for the same effective oxide thickness). This may affect the device characteristics and performance significantly and hence it becomes important to recognize this effect especially for high-K gate dielectric SOI MOSFETs. Our model can be easily implemented in a circuit simulator to include this effect.

**Figure Captions**

Figure 1        Cross-sectional view of SOI MOSFET showing internal parasitic fringe capacitance.

Figure 2        Fringing field from the bottom of gate electrode to drain (or source).

Figure 3        Potential due to a uniformly charged plate.

Figure 4        Internal fringe capacitance variation with gate oxide permittivity for (a) Lg = 60 nm, (b) Lg = 40 nm. Gate width W is fixed at 1 $\mu$m, EOT = 2.0 nm, $\varepsilon_{sp}$ = 3.9.

Figure 5        Calculated surface potential variation along the channel for $\varepsilon_{ox}$ = 60 with and without the effect of fringe capacitance. The parameters used are: $V_D$ = 0.05 V, $V_G$ = 0.02 V, EOT = 2.0 nm, $\varepsilon_{sp}$ = 3.9, $N_A$ = 1x10$^{16}$ cm$^{-3}$, W= 1 $\mu$m.

Figure 6        Comparison of the simulated and calculated threshold voltage variation versus gate dielectric constant for (a) Without S/D-Gate overlap (b) With S/D-Gate overlap.   The parameters used are:  $V_D$ = 0.05 V, EOT = 2.0 nm, $\varepsilon_{sp}$ = 3.9, $N_A$ = 1x10$^{16}$ cm$^{-3}$, W= 1 $\mu$m.

Figure 7        Effect of including Quantum Mechanical effects in simulation. The parameters used are: $V_D$ = 0.05 V, EOT = 2.0 nm, $\varepsilon_{sp}$ = 3.9, $N_A$ = 1x10$^{16}$ cm$^{-3}$, W= 1 $\mu$m.

Table 1: Device parameters used in the simulation

| Parameter | Value |
|---|---|
| Source/Drain doping | $2x10^{20}$ cm$^{-3}$ |
| Channel doping | $1x10^{16}$ cm$^{-3}$ |
| Effective Oxide Thickness (EOT)$^{*}$ | 2.0 nm |
| Work Function of gate material | 4.5 V |
| Silicon film thickness | 15 nm |
| Spacer oxide thickness | 25 nm |
| BOX thickness | 100 nm |
| Substrate Thickness | 100 nm |
| Gate electrode thickness | 25 nm |
| Source/drain – Gate overlap | 5 nm |

$^{*}$(Physical oxide thickness = EOT$\times\varepsilon_{ox}/\varepsilon_{SiO2}$)

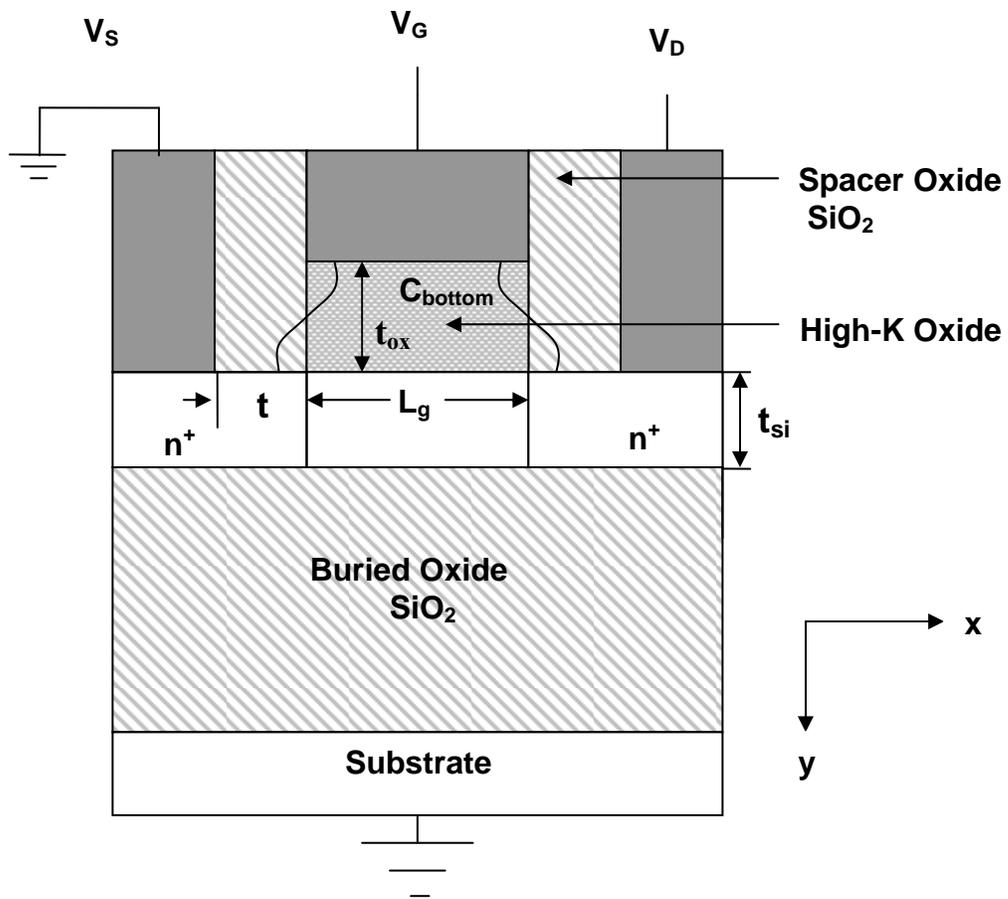

**Figure 1**

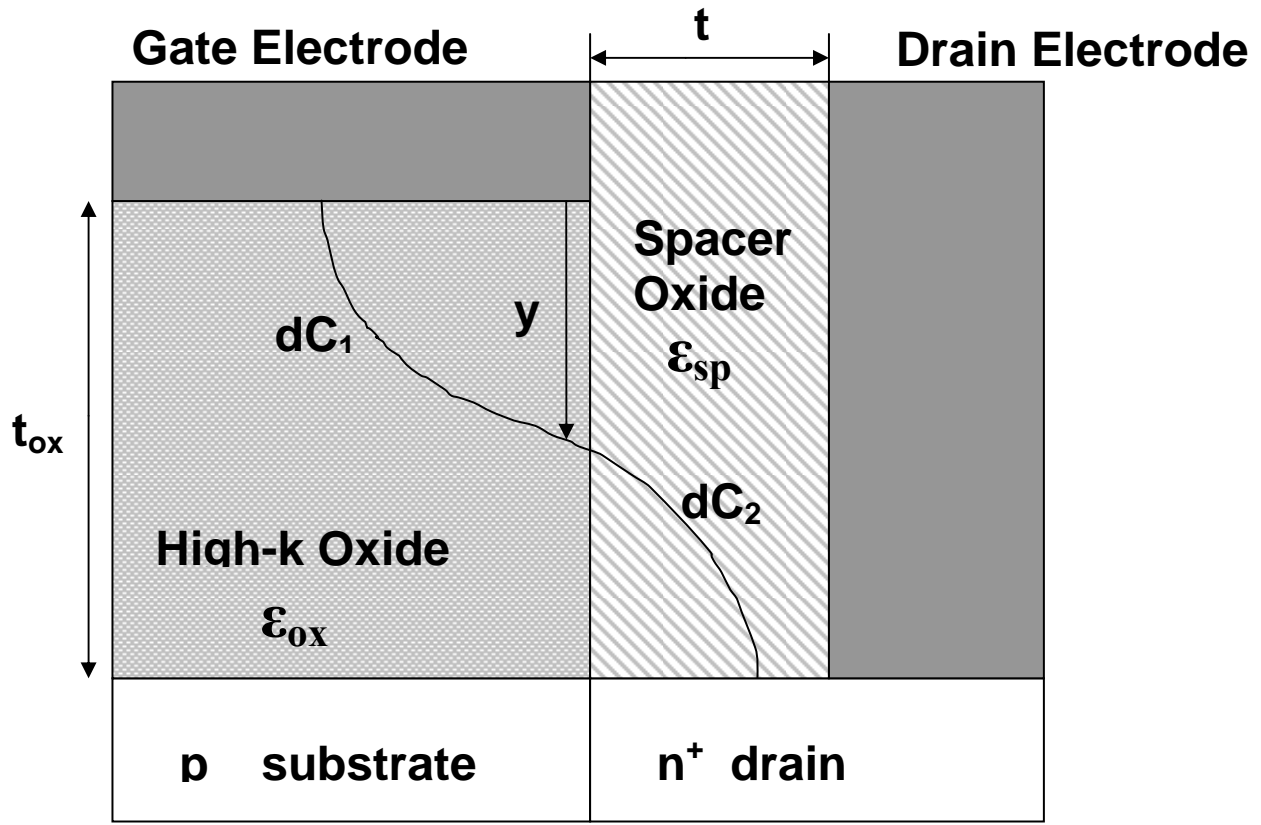

**Figure 2**

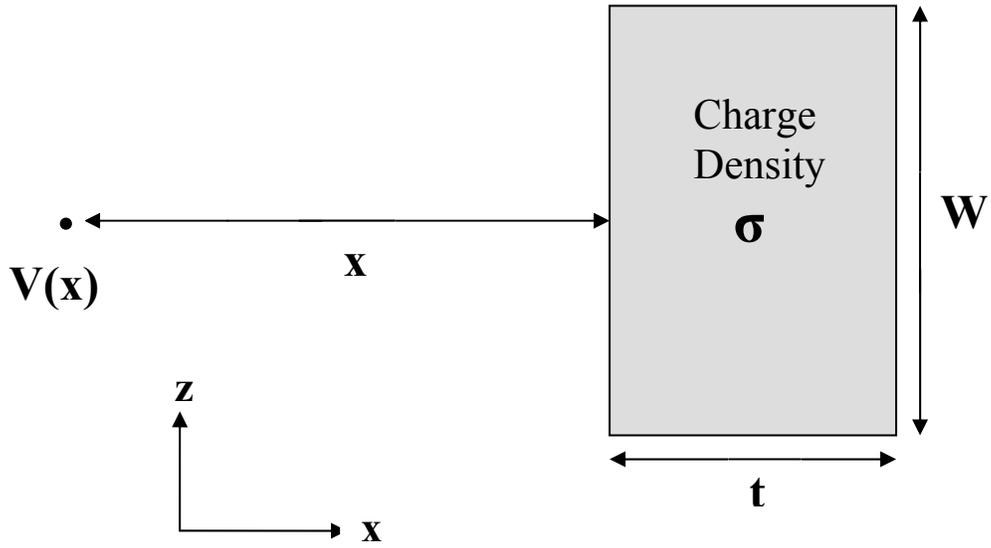

**Figure 3**

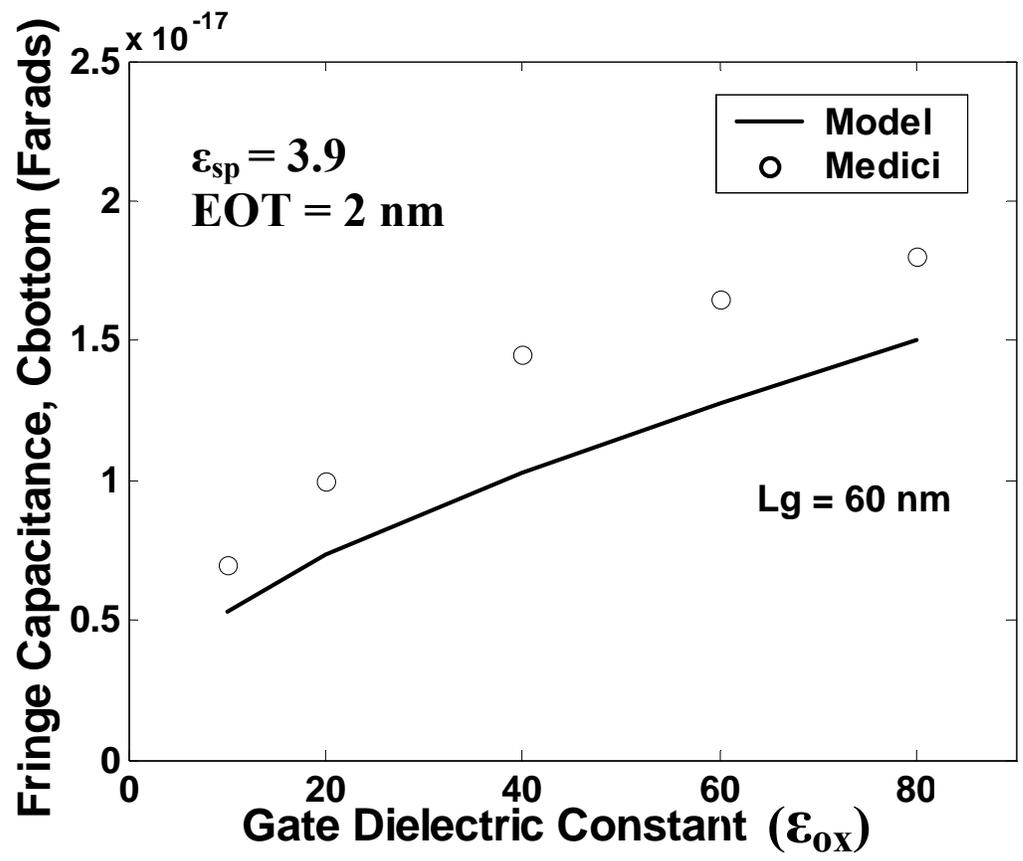

**Figure 4(a)**

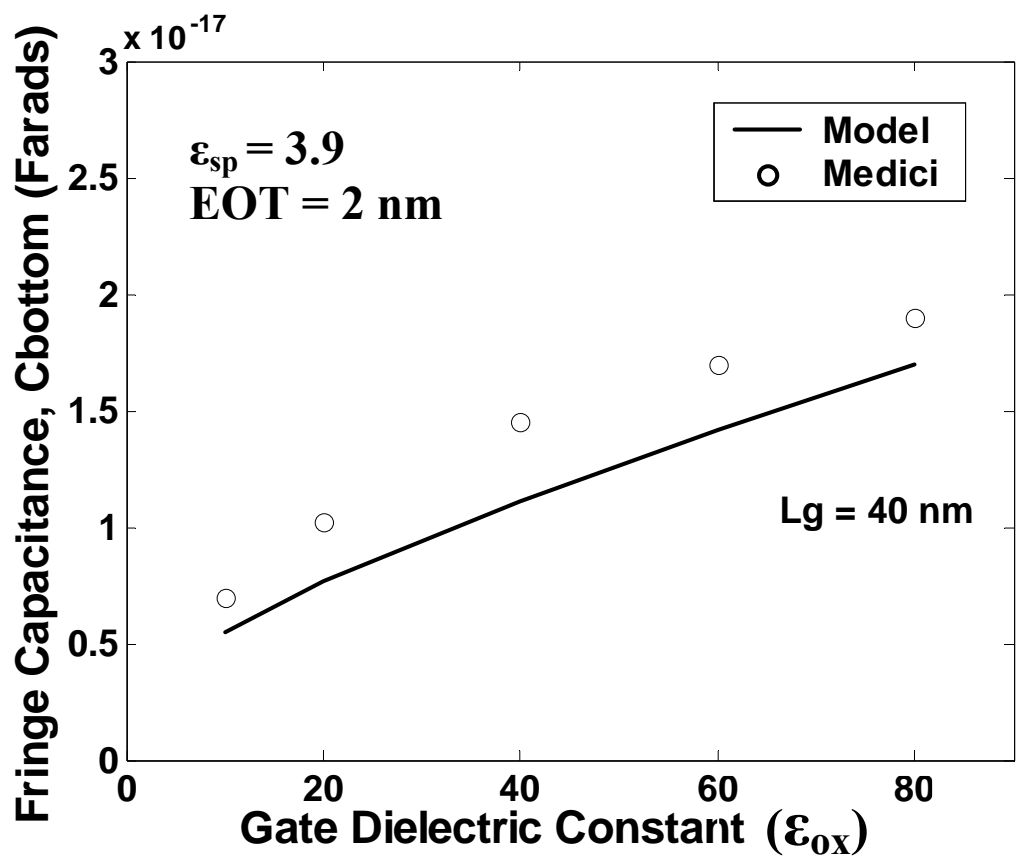

**Figure 4(b)**

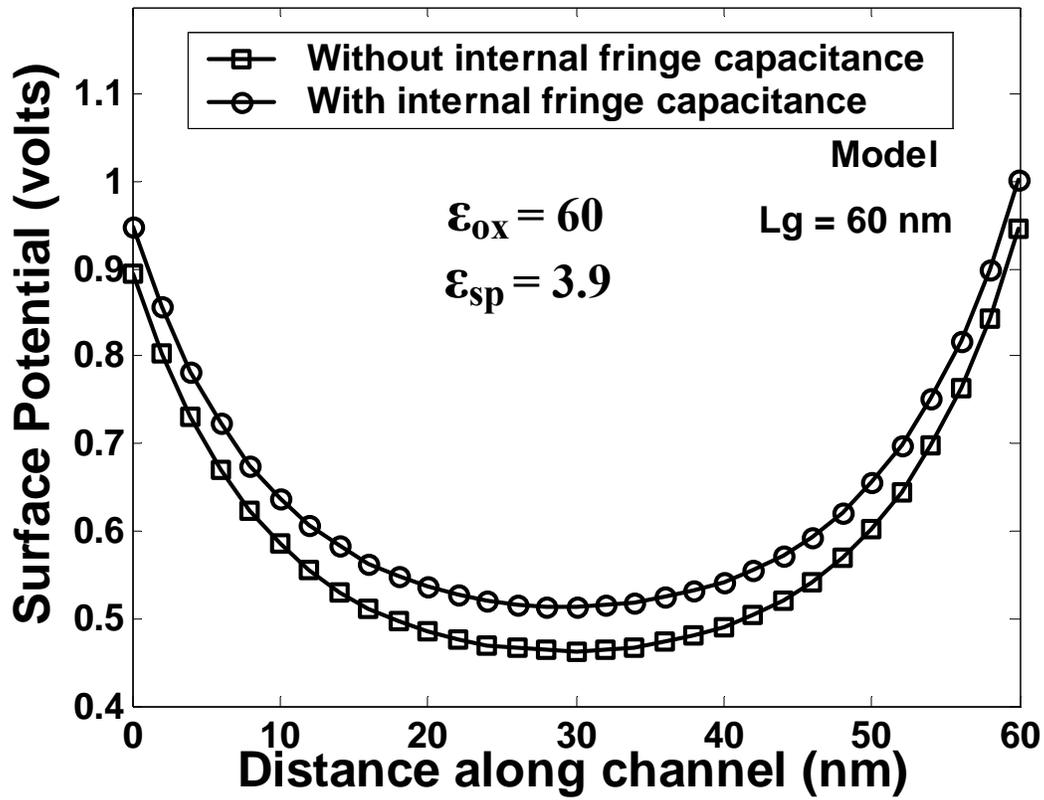

**Figure 5**

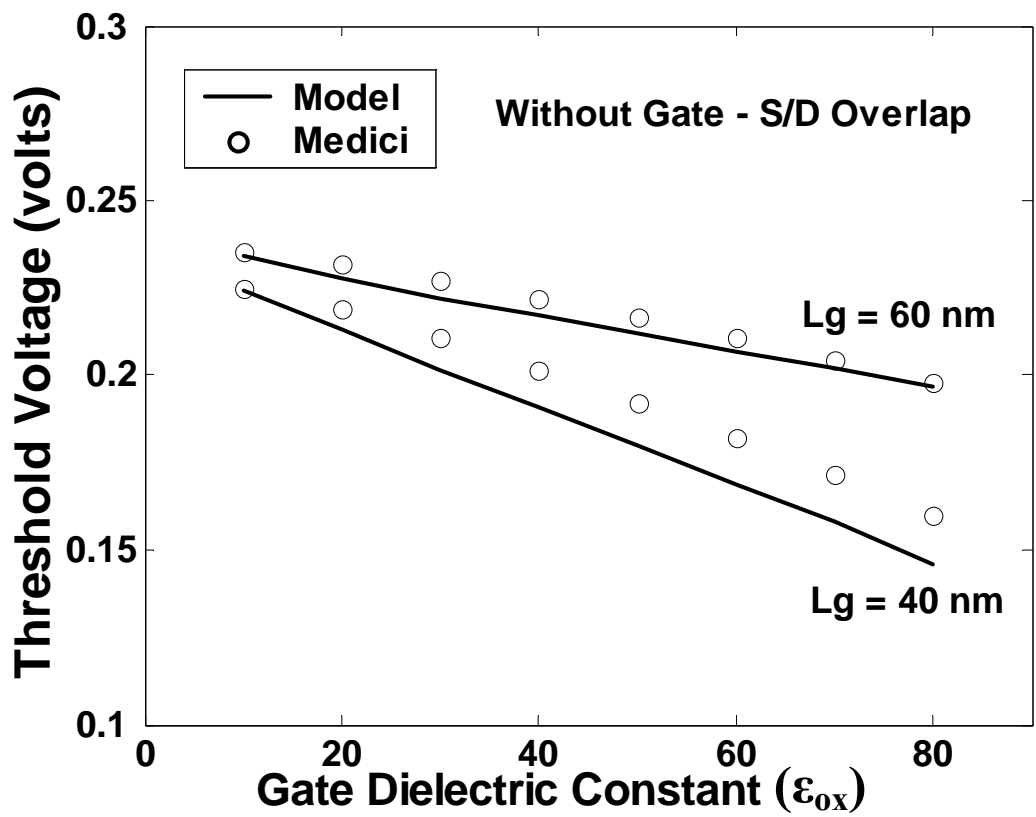

**Figure 6(a)**

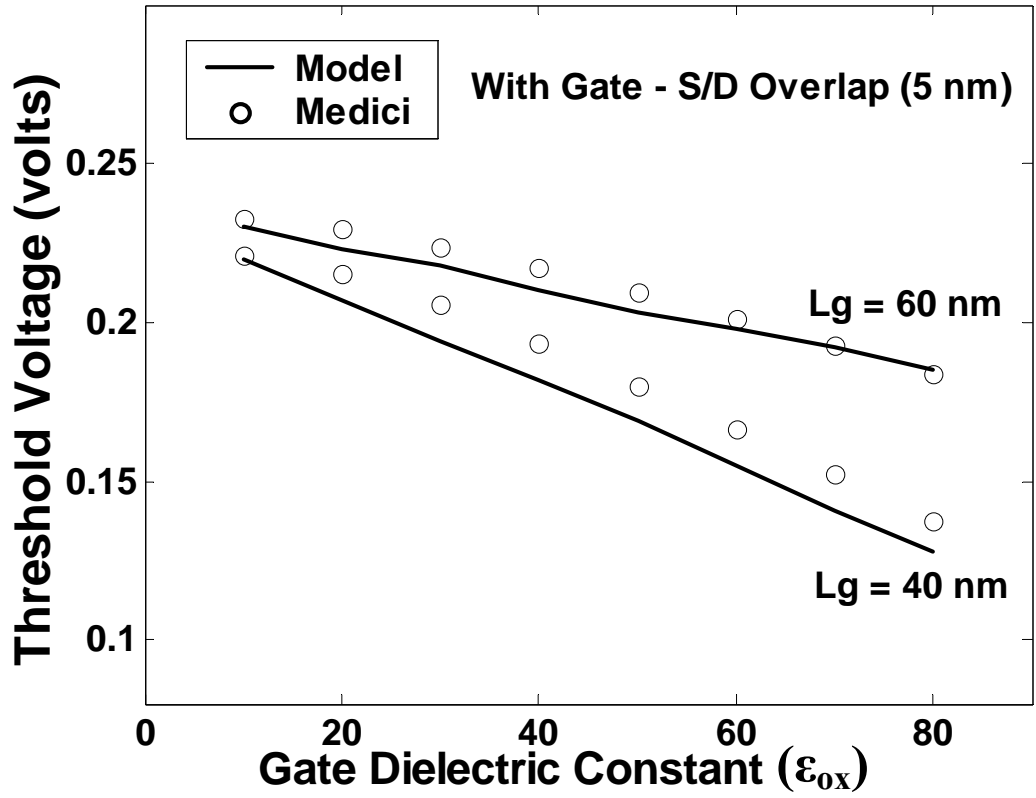

**Figure 6(b)**

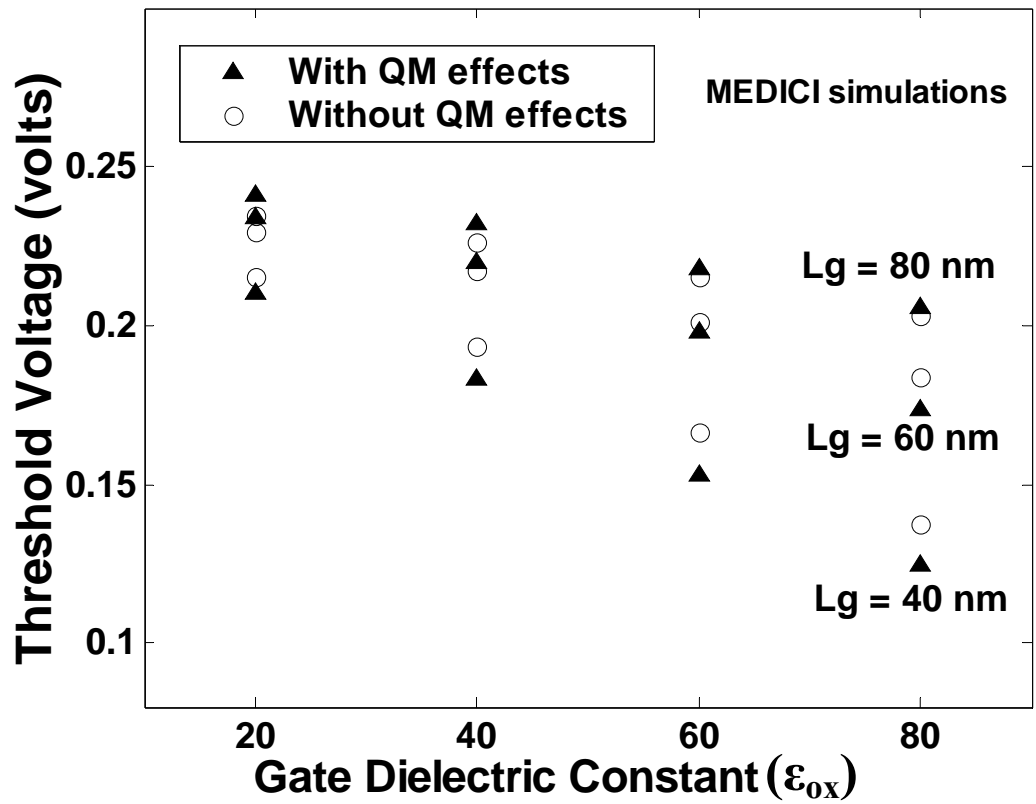

**Figure 7**

# Author Biographies

**M. Jagadesh Kumar** (SM'99) was born in Mamidala, Nalgonda District, Andhra Pradesh, India. He received the M.S. and Ph.D. degrees in electrical engineering from the Indian Institute of Technology, Madras, India. From 1991 to 1994, he performed post-doctoral research in modeling and processing of high-speed bipolar transistors with the Department of Electrical and Computer Engineering, University of Waterloo, Waterloo, ON, Canada. While with the University of Waterloo, he also did research on amorphous silicon TFTs. From July 1994 to December 1995, he was initially with the Department of Electronics and Electrical Communication Engineering, Indian Institute of Technology, Kharagpur, India, and then joined the Department of Electrical Engineering, Indian Institute of Technology, Delhi, India, where he became an Associate Professor in July 1997 and a Full Professor in January 2005. His teaching has often been rated as outstanding by the Faculty Appraisal Committee, IIT, Delhi. His research interests are in VLSI device modeling and simulation for nanoscale applications, integrated-circuit technology, and power semiconductor devices. He has more than 100 publications in peer reviewed journals and conferences.

Dr. Kumar is a Fellow of Institute of Electronics and Telecommunication Engineers (IETE),India. He has reviewed extensively for different journals including *IEEE Trans. on Electron Devices, IEE Proc. on Circuits, Devices and Systems, Electronics Letters and Solid-state Electronics.* He was Chairman, Fellowship Committee, *The Sixteenth International Conference on VLSI Design,* January 4-8, 2003, New Delhi, India. He was Chairman of the Technical Committee for High Frequency Devices, *12th International Workshop on the Physics of Semiconductor Devices*, December 13-17, 2005, New Delhi, India.

**Vivek Venkataraman,** *Student Member, IEEE,* is currently pursuing his B.Tech. degree in electrical engineering from Indian Institute of Technology, Delhi, India. Device modeling and simulation for nanoscale applications is one of his research interests.

**Sumeet Kumar Gupta,** *Student Member, IEEE,* is currently pursuing his B.Tech. degree in electrical engineering from Indian Institute of Technology, Delhi, India. Device modeling and simulation for nanoscale applications is one of his research interests.